\documentclass[conference,a4paper]{IEEEtran}
\usepackage{etoolbox}
\makeatletter
\patchcmd{\@makecaption}
  {\scshape}
  {}
  {}
  {}
\makeatother
\makeatletter
\patchcmd{\@makecaption}
  {\\}
  {.\ }
  {}
  {}
\makeatother
\def\tablename{Table}
\IEEEoverridecommandlockouts

\usepackage{cite}
\usepackage{amsmath,amssymb,amsfonts}
\usepackage{algorithmic}
\usepackage{graphicx}
\usepackage[colorlinks,linkcolor=blue]{hyperref}
\usepackage{floatrow}
\usepackage{booktabs}
\usepackage{textcomp}
\usepackage{xcolor}
\def\BibTeX{{\rm B\kern-.05em{\sc i\kern-.025em b}\kern-.08em
    T\kern-.1667em\lower.7ex\hbox{E}\kern-.125emX}}
\begin{document}

\title{BigWavGAN: A Wave-To-Wave Generative Adversarial Network for Music Super-Resolution\\
}

\author{\IEEEauthorblockN{Yenan Zhang}
\IEEEauthorblockA{\textit{Graduate School of Fundamental Science and Engineering}\\
\textit{Waseda University}\\
Tokyo, Japan}
\and
\IEEEauthorblockN{Hiroshi Watanabe}
\IEEEauthorblockA{\textit{Graduate School of Fundamental Science and Engineering} \\
\textit{Waseda University}\\
Tokyo, Japan}
}

\maketitle

\begin{abstract}
Generally, Deep Neural Networks (DNNs) are expected to have high performance when their model size is large. However, large models failed to produce high-quality results commensurate with their scale in music Super-Resolution (SR). We attribute this to that DNNs cannot learn information commensurate with their size from standard mean square error losses. To unleash the potential of large DNN models in music SR, we propose BigWavGAN, which incorporates Demucs, a large-scale wave-to-wave model, with State-Of-The-Art (SOTA) discriminators and adversarial training strategies. Our discriminator consists of Multi-Scale Discriminator (MSD) and Multi-Resolution Discriminator (MRD). During inference, since only the generator is utilized, there are no additional parameters or computational resources required compared to the baseline model Demucs. Objective evaluation affirms the effectiveness of BigWavGAN in music SR. Subjective evaluations indicate that BigWavGAN can generate music with significantly high perceptual quality over the baseline model. Notably, BigWavGAN surpasses the SOTA music SR model in both simulated and real-world scenarios. Moreover, BigWavGAN represents its superior generalization ability to address out-of-distribution data. The conducted ablation study reveals the importance of our discriminators and training strategies. Samples are available on the demo page: https://mannmaruko.github.io/demopage/BigWavGAN/d.html.
\end{abstract}


\section{Introduction}\label{introduction}
Audio Super-Resolution (SR) involves the transformation of low-resolution (\textit{i.e.}, narrow-band) input into high-resolution (\textit{i.e.}, wide-band) audio, which gives the low-resolution audio more details and a brighter tone. This paper specifically delves into the task of music SR, which is challenging due to the broad frequency response in music. In this paper, we concentrate on the music SR task of solo piano.

Deep Neural Networks (DNNs) are often associated with high performance when the model size is large. However, the research by Zhang \textit{et\;al.} \cite{zhang2023PhaseRepair} indicated that the large-scale model cannot generate music with the quality that is commensurate with its model size, mainly due to phase distortion. We attribute this to that models cannot learn information (\textit{e.g.}, correct phase information) commensurate with their size through standard Mean Square Error (MSE) losses. In recent years, several works utilized neural vocoders to address audio SR tasks. Vocoders map mel-spectrograms to raw waveforms \cite{hu2020PhaseAwareinterspeech,liu2022neuralvocoderisallyouneed,zhang2023PhaseRepair}. Notably, a State-Of-The-Art (SOTA) neural vocoder named BigVGAN is characterized by a large-size generator with up to 112M parameters \cite{lee2022bigvgan}. BigVGAN can synthesize high-fidelity audio and shows its superior zero-shot performance across various out-of-distribution scenarios. However, in the task of audio SR, there is no wave-to-wave GAN-based model in such a large model size. This inspired us to explore the large-scale wave-to-wave GAN model in music SR with high performance and superior generalization ability.

To this end, we propose BigWavGAN, which integrates Demucs, a large-scale wave-to-wave model containing 134M parameters, with SOTA discriminators and adversarial training strategies. Precisely, the combination of Multi-Scale Discriminator (MSD) and Multi-Resolution Discriminator (MRD) constitutes the discriminator of BigWavGAN. During inference, only the generator is utilized, resulting in no
additional parameters or computational requirements compared
to the baseline model Demucs. We evaluate BigWavGAN from both objective and subjective perspectives: (1) The objective evaluation affirms the effectiveness of BigWavGAN in music SR. (2) The subjective evaluations indicate that the proposed BigWavGAN is capable of generating high-resolution music with better perceptual quality than its baseline. (3) Moreover, BigWavGAN represents its strong ability to handle out-of-distribution data. (4) Notably, BigWavGAN surpasses the SOTA music SR model in both simulated and real-world scenarios. (5) At last, the ablation study unveils the importance of our discriminators and training strategies. In conclusion, BigWavGAN successfully unleashes the potential in the baseline model without additional computation or parameters. 

\section{Related work}
Demucs is a large-scale model initially designed for music source separation \cite{defossez2019demucs} but also generated fairly good results in other tasks, such as music SR \cite{zhang2023PhaseRepair} and music enhancement \cite{2022MusicEnhancement}. Due to the large size of Demucs, it was anticipated to produce high-quality results in music SR. However, Demucs still yielded results with annoying artifacts \cite{zhang2023PhaseRepair}. Zhang \textit{et\;al.} \cite{zhang2023PhaseRepair} also represented that besides Demucs, AudioUNet \cite{kuleshov2017AudioUnet} and SEANet \cite{tagliasacchi2020seanet} cannot generate high-quality audio due to phase problem when trained by standard MSE losses. To address the artifacts, Zhang \textit{et\;al.} employed a neural vocoder to rectify the distorted phase generated by Demucs. Nevertheless, the improvements brought by phase repair remain limited, which indicates that introducing adversarial training into the model can lead the model to learn more information.  

Recent publications have delved into GAN-based models in audio SR. Compared to models trained with standard MSE losses, GAN-based models exhibit a superior capability to generate results with better perceptual quality \cite{moliner2022BEHM}. BEHMGAN is the SOTA of GAN-based music SR model. It comprises a complex U-net as the generator and the MSD discriminator from MelGAN \cite{kumar2019melgan}.  MelGAN is the first work that successfully synthesizes realistic speeches by training GANs without additional distillation or perceptual loss functions. TFGAN is a lightweight vocoder for speech, which employs MSD and a single-resolution frequency discriminator as its discriminator \cite{tian2020tfgan}. TFGAN has been used in audio SR \cite{liu2022neuralvocoderisallyouneed,zhang2023PhaseRepair}. Jiang \textit{et\;al.} proposed an advanced neural vocoder named UnivNet, in which MRD was proposed and was proved to effectively improve the performance of MelGAN \cite{jang2021univnet}. Lee \textit{et\;al.} proposed BigVGAN, which uses SOTA adversarial training strategies at an unprecedented scale of more than 100M parameters \cite{lee2022bigvgan}.

\section{Proposed method}
\floatsetup[table]{capposition=top}
\begin{table*}[h]
\caption{LSD scores on the MAESTRO dataset. The bold represents the top two LSD scores.}
\begin{center}
\resizebox{14cm}{!}{
\begin{tabular}{l|cc|cccc|c}
\toprule[2pt]
& MSD & MRD & 2.5 kHz & 3.0 kHz &3.5 kHz & 4.0 kHz & AVG LSD \\
\midrule
Input & - & - & 2.43 & 2.19 & 1.97 & 1.78 & 2.09 \\
\midrule
BEHMGAN \cite{moliner2022BEHM} & \checkmark & - & 1.89 & 1.01 & 1.79 & 1.77 & 1.61 \\
\midrule
BigWavGAN (proposed) & \checkmark & \checkmark & \textbf{0.83} & \textbf{0.79} & \textbf{0.76} & \textbf{0.73} & \textbf{0.78} \\
  \quad - w/o MRD & \checkmark & - & 0.93 & 0.88 & 0.82 & \textbf{0.73} & 0.84 \\
  \quad \quad - w/o MSD (Demucs) & - & - & \textbf{0.82} & \textbf{0.74} & \textbf{0.68} & \textbf{0.64} & \textbf{0.72} \\
\bottomrule[2pt]
\end{tabular}
}
\end{center}
\label{table:lsd of the ablation study}
\end{table*}

Although Demucs is a large-scale model with 134M parameters, it didn't generate high-quality waveforms commensurate with its large size in music SR \cite{zhang2023PhaseRepair}. To unleash the potential of Demucs, we propose BigWavGAN for wave-to-wave music SR, which incorporates Demucs with SOTA discriminators and adversarial training strategies. 
\subsection{Architecture of BigWavGAN}
\begin{figure}[t] 
\centering
\includegraphics[width=\linewidth]{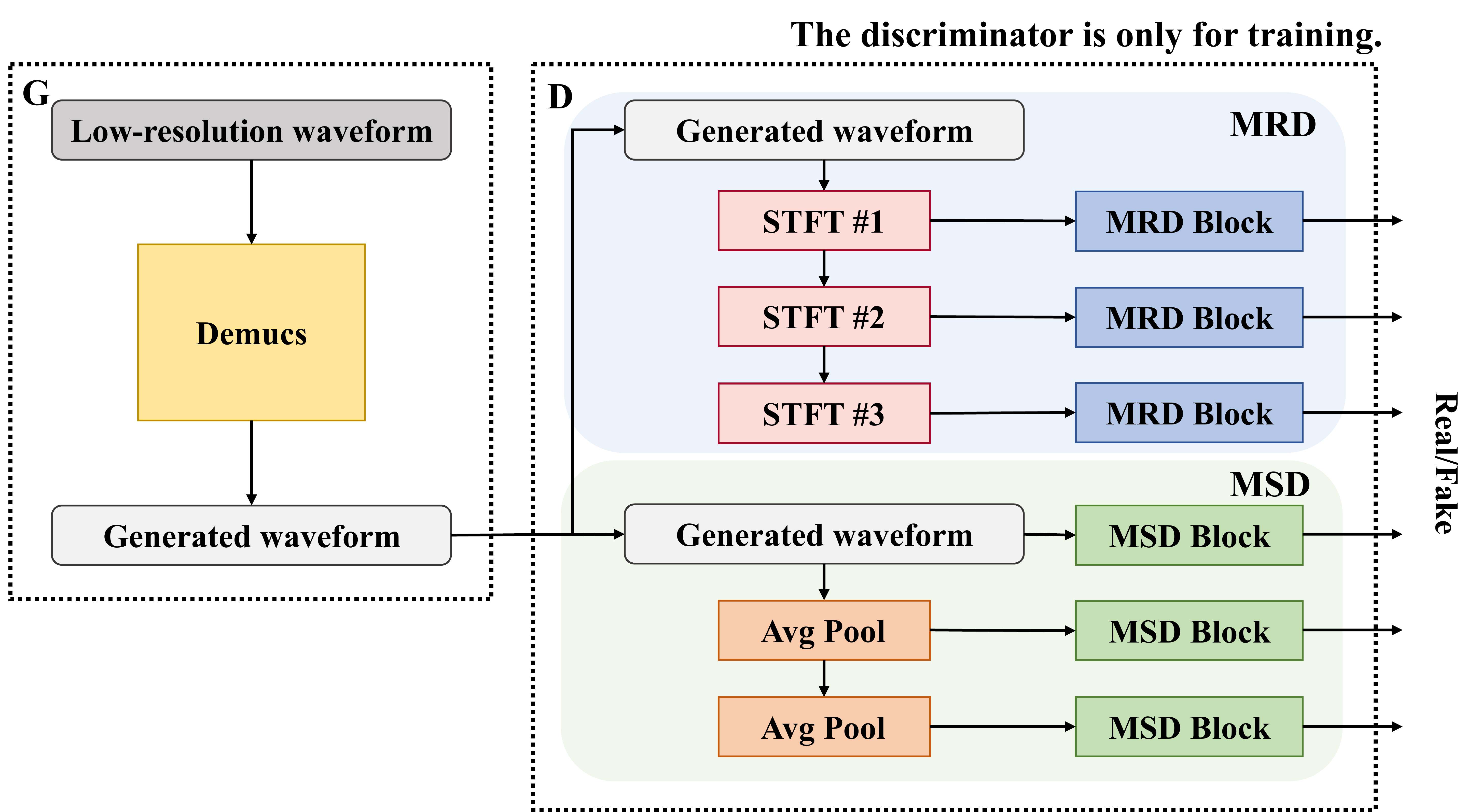}
\caption{Overview of the architecture of BigWavGAN.}
\label{fig:overview}
\end{figure}

The overview of BigWavGAN's architecture is shown in Fig. \ref{fig:overview}. The generator of BigWavGAN has the identical architecture with Demcus from \cite{defossez2019demucs}. It is a wave domain U-net model leveraging a Long Short-Term Memory (LSTM) recurrent neural network layer as the bottleneck. 

BigWavGAN benefits from the two types of discriminators: MSD and MRD. MSD works in the time domain, where each sub-discriminator receives down-sampled 1-D waveform signals at downsampling ratios of 1, 2, and 4. MRD works in the frequency domain, which also comprises several sub-discriminators operating on multiple 2-D spectrograms with different Short-Time Fourier Transform (STFT) resolutions. On top of standard MSE losses, applying different types of discriminators to cross domains (\textit{i.e.}, time and frequency domains) guides the generator to restore high-resolution music that is realistic in multiple domains and resolutions, minimizing annoying artifacts that are common for wave-to-wave models. 

Our choice of MRD with MSD is not common in related vocoder publications, in which the  Multi-Period Discriminator (MPD) is widely used \cite{kong2020hifi,jang2021univnet,lee2022bigvgan}. However, since MPD reshapes the 1-D waveform into 2-D matrices at multiple periods, it requires much more computational resources than MSD, making the training difficult for low-resource environments. 

To improve training efficiency, we decided to replace MPD by MSD. Although the design of MPD and MSD is different, they all work in the time domain, which implies that MSD could be an alternative to MPD in order to similarly capture details in the waveform. The evaluation results in section \ref{sec:eval} reveal BigWavGAN's superior performance, validating the success of combining MSD and MRD as the discriminator. 

Adversarial training of large-scale models tend to be unstable. To stabilize the training, we utilized the training strategies of BigVGAN. Lee \textit{et\;al.} \cite{lee2022bigvgan} made lots of efforts on maintaining the stability of large-scale GAN training and the high-speed practical usability. We believe that these training strategies are suitable for training non-vocoder models with a similar scale, and introduced these strategies into BigWavGAN's training to ensure training stability.

\subsection{Training Objectives}
In terms of training objectives, we applied $\textit{L}_\textit{G}$ for generator and $\textit{L}_\textit{D}$ for discriminator, respectively:

\begin{equation}
    L_G = \sum_{k=1}^K \bigg[ L_{adv}\left(G; D_k\right) + \lambda_{fm}L_{fm} \left(G; D_k\right) \bigg] + \lambda_{mel}L_{mel} \left(G\right),
\end{equation}

\begin{equation}
L_D = \sum_{k=1}^K \bigg[ L_{adv}\left( D_k;G\right) \bigg],
\end{equation}
where $K=3$, $D_k$ denotes the $k$-th MSD or MRD submodules. $L_{adv}$ stands for adversarial losses, $L_{fm}$ stands for feature matching losses, $L_{mel}$ stands for mel losses. We used the scalar weights $\lambda_{fm} = 2$ and $\lambda_{mel} = 45$ identically as \cite{lee2022bigvgan}.

$L_{adv}$ uses the least-square GAN as follows:

\begin{equation}
    L_{adv}\left(G;D_k\right) = \mathbb{E}_s \bigg[ (D_k(G(s))-1)^2\bigg],
\end{equation}

\begin{equation}
    L_{adv}\left(D_k;G\right) = \mathbb{E}_{(x,s)} \bigg[ (D_k(x)-1)^2 + (D_k(G(s)))^2\bigg],
\end{equation}
where $s$ is the input low-resolution waveform, $x$ is the ground-truth waveform. 

The feature matching loss $L_{fm}$ minimizes the $l_1$ distance for every intermediate features from the discriminator layers: 
\begin{equation}
    L_{fm}(G;D_k) = \mathbb{E}_{(x,s)} \bigg[ \sum_{i=1}^T \frac{1}{N} || D_k^i(x) - D_k^i(G(s)) ||_1 \bigg],
\end{equation}
where $T$ is the number of layers of the sub-discriminator $D_k$. 

The generator loss $L_G$ also has the spectral $l_1$ regression loss between the mel spectrogram of the synthesized waveform and the corresponding ground-truth:
\begin{equation}
    L_{mel}(G) = \mathbb{E}_{(x,s)} \bigg[ || \phi(x) - \phi(G(s)) ||_1 \bigg],
\end{equation}
where $\phi$ is the STFT with mel filter bank that converts the waveform into mel-spectrogram.

\section{Experiments}
We used the MAESTRO dataset \cite{Hawthorne2018MAESTRO} for training. We simulated the low-resolution music by means of following \cite{zhang2023PhaseRepair, liu2022neuralvocoderisallyouneed}. To handle real-world low-resolution music recordings which have various bandwidths, we simulated the input bandwidth ranging from 2.0 kHz to 4.0 kHz on the fly during training. The models involved in our evaluation all work at the sampling rate of 16 kHz with a target bandwidth of 8 kHz, except BEHMGAN. The configurations of the low-pass filters used to simulate low-resolution audio are identical to that in \cite{zhang2023PhaseRepair}. Hereby, the proposed BigWavGAN can deal with any bandwidths between 2.0 kHz and 4.0 kHz. 

The implementation of BigWavGAN's generator (\textit{i.e.}, Demucs) is from \cite{defossez2019demucs}. We implemented MSD and MRD by utilizing the open-source code from \cite{lee2022bigvgan} and \cite{kong2020hifi} respectively. During training, the batch size is 10, each music segment is 2.56 seconds long. We trained BigWavGAN for 1M iterations with the same training strategies as BigVGAN \cite{lee2022bigvgan}. As BigVGAN is similar to our BigWavGAN in model size, keeping the same training strategy contributed to the stable training of BigWavGAN. 

For the baseline Demucs, we used the checkpoint from \cite{zhang2023PhaseRepair}. We used the official checkpoints of BEHMGAN \cite{moliner2022BEHM} for comparison. Music generated by BEHMGAN were resampled from 22.05 kHz to 16 kHz for a fair evaluation. Furthermore, in order to explore the effectiveness of the discriminator and training strategies, we also trained a model denoted as BigWavGAN w/o MRD which is trained by discriminators and strategies from TFGAN \cite{tian2020tfgan}. TFGAN combines MSD with a single-resolution frequency discriminator instead of MRD. We implemented this training by using an unofficial implementation\footnote{https://github.com/rishikksh20/TFGAN
} and trained this model for 1M iterations.



\section{Evaluation}
\label{sec:eval}
We evaluated the proposed BigWavGAN from both objective and subjective perspectives. 
\begin{figure*}[h] 
\centering
\includegraphics[width=\linewidth]{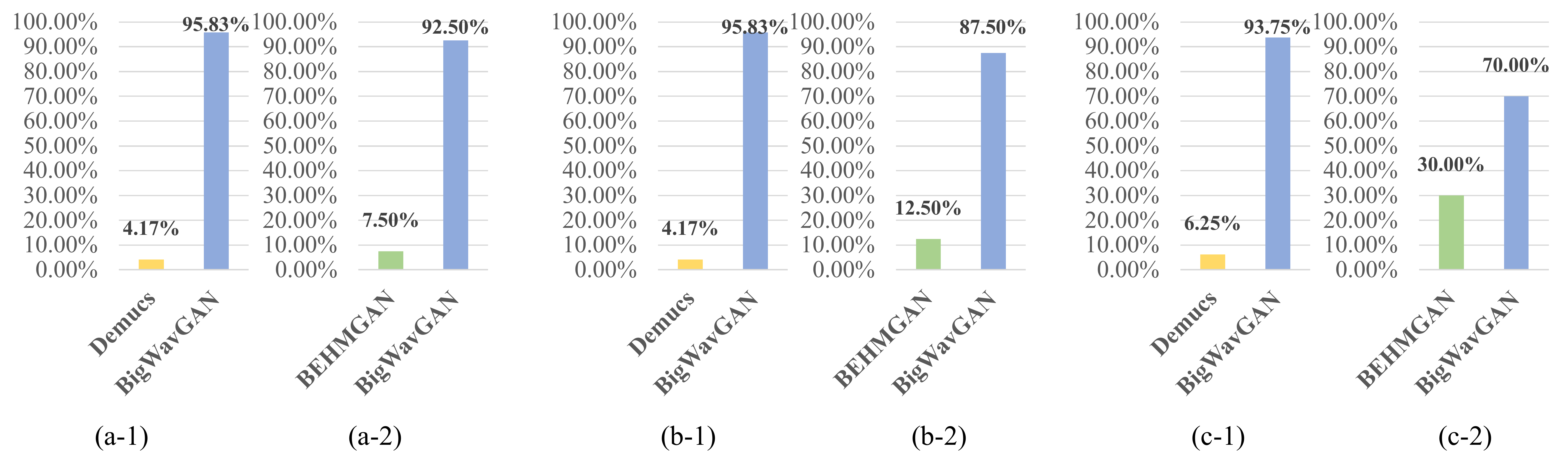}
\caption{The results of A/B listening tests: (a) is tested on MAESTRO; (b) is tested on MusicNet; (c) is tested on denoised real historical recordings.}
\label{fig:abtests}
\end{figure*}
\subsection{Objective Evaluation}
We used Log-Spectral Distance (LSD) as the objective metric, which is widely used in audio SR tasks \cite{liu2022neuralvocoderisallyouneed, moliner2022BEHM}. We calculated the LSD scores at four representative bandwidths (\textit{i.e.}, 2.5 kHz, 3.0 kHz, 3.5 kHz, 4.0 kHz). The results of LSD are illustrated in Tab. \ref{table:lsd of the ablation study}. Note that the proposed BigWavGAN can handle any bandwidth from 2.0 kHz to 4.0 kHz. 

In terms of LSD scores, the four models all successfully achieved music SR since all the generated results received much better LSD scores than low-resolution inputs. BEHMGAN was trained on inputs with bandwidths around 3.0 kHz, as 3.0 kHz was believed to be the typical bandwidth of real historical recordings \cite{moliner2022BEHM}. Consequently, BEHMGAN performed well at 3.0 kHz. Nevertheless, the proposed BigWavGAN still outperformed BEHMGAN at this bandwidth. 

In order to explore the importance of the discriminator and training strategies, we compared BigWavGAN with a variant that has only a single-resolution frequency discriminator combined with the MSD, \textit{i.e.}, BigWavGAN w/o MRD. We found that the proposed BigWavGAN, which utilizes MRD and MSD with the training strategies from \cite{kong2020hifi, lee2022bigvgan}, outperformed the ``w/o MRD'' variant overall. Since LSD is a metric working on the frequency domain, compared with the single-resolution frequency discriminator, the multi-resolution frequency discriminator (MRD) seems to have improved the LSD score by forcing the model to concentrate more on the fidelity of music in the frequency domain.

The proposed BigWavGAN acquired a slightly worse LSD score than the baseline model Demucs. We consider this difference in LSD score as the result of the common phenomenon that objective metrics tend to give generative methods lower scores than their non-generative counterparts \cite{liu2022neuralvocoderisallyouneed, zhang2023PhaseRepair}. This can be explained as that generative models tend to generate results similar rather than exactly identical to ground truth. To show that BigWavGAN can restore music with better perceptual quality, subjective evaluations were conducted. 
\subsection{Subjective Evaluation}
Although LSD can well reflect how well the high frequency in the magnitude is recovered, it cannot reflect the degree of the artifacts and has been observed not to correlate with perceptual audio quality \cite{liu2022neuralvocoderisallyouneed,zhang2023PhaseRepair}. To this end, we conducted a set of subjective evaluations to identify the advantage of the proposed BigWavGAN. The subjective evaluation is in the style of A/B test, rather than mean opinion score test, because A/B test can better measure tiny differences between two models. Since A/B test cannot handle multiple models at once, we conduct multiple A/B tests (\textit{e.g.}, BigWavGAN vs Demucs, BigWavGAN vs BEHMGAN) for a more comprehensive analysis. 

We conducted A/B tests on 3 different datasets: (a) four tracks from MAESTRO \cite{Hawthorne2018MAESTRO}, (b) four piano tracks from MusicNet \cite{thickstun2016learningmusicnet}, and (c) for real historical piano recordings provided in \cite{moliner2022BEHM}. We only trained our BigWavGAN on MAESTRO dataset, then applied it to out-of-distribution data (\textit{i.e.}, MusicNet, and real-world historical recordings) in the zero-shot condition to evaluate its generalization ability. To avoid too many testing samples and the consequent auditory fatigue in participants, the input bandwidth is set to 3.0 kHz, except for the real-world recordings.  We selected the above 12 tracks to cover different musicians, periods, and styles. The duration of each music clip has been standardized to 10 seconds, and all audio clips have a normalized loudness for accurate evaluation. Twelve and eleven people with no background in audio engineering participated our subjective tests for BigWavGAN vs Demucs and BigWavGAN vs BEHMGAN respectively

The results of subjective evaluations are illustrated in Fig \ref{fig:abtests}. In all three datasets, BigWavGAN significantly improved Demucs in terms of perceptual quality by a large margin. This also reveals that BigWavGAN achieved superior generalization to out-of-distribution data. Similar advantages of BigWavGAN are observed when it is compared with BEHMGAN, the SOTA music SR model. We further analyze the trends in the preference of BigWavGAN and BEHMGAN. First, in Fig. \ref{fig:abtests} we can see that BEHMGAN's preference increased from MAESTRO (a-2) to MusicNet (b-2), \textit{i.e.}, the preferences changed from 7.50\% vs 92.50\% to 12.50\% vs 87.50\%. This is because BEHMGAN was trained on MusicNet. Although MusicNet is out-of-distribution data to our BigWavGAN, we outperformed BEHMGAN by a large margin, validating the strong generalization of BigWavGAN again.

Although in the real-world historical recordings, the preference of BEHMGAN further increased to 30.00\% vs 70.00\%, BigWavGAN still outperformed BEHMGAN with a large margin. We think the less advantage of BigWavGAN in real-world condition is due to our limited simulation in training data. In the future, we would like to explore more realistic simulation techniques and further improve our model's performance in real-world historical recordings. 

\section{Conclusion}
In this paper, we proposed a large-scale wave-to-wave model called BigWavGAN for music Super-Resolution (SR). The model integrates a large-size generator (\textit{i.e.}, Demucs with up to 134M parameters), with the State-Of-The-Art (SOTA) discriminators and adversarial training strategies. The discriminator of the proposed BigWavGAN consists of Multi-Scale Discriminator (MSD) and Multi-Resolution Discriminator (MRD). During inference phase since only the generator will be used, there are no additional parameters or computational resources required during inference compared to the baseline model Demucs. We evaluated BigWavGAN from both objective and subjective perspectives. The objective evaluation indicates the effectiveness of BigWavGAN in music SR. The results of a set of subjective evaluations demonstrate that BigWavGAN can produce high-resolution music in significantly better perceptual quality compared to the baseline model Demucs. Notably, the subjective evaluations also indicate that BigWavGAN surpasses the SOTA music SR model in both simulated and real-world scenarios. Moreover, it also implies that BigWavGAN achieves superior generalization ability to address out-of-distribution data including real historical recordings. Therefore, BigWavGAN successfully unleashes the potential of the large-scale Demucs in music SR. In the future, we hope to explore more realistic simulation techniques to further improve BigWavGAN in real-world scenarios, as well as to further extend BigWavGAN to more tasks.


\vspace{12pt}

\bibliographystyle{unsrt}
\bibliography{MyReferences}

\end{document}